\documentclass[runningheads]{llncs}
\usepackage{graphicx}

\usepackage{booktabs}
\usepackage{adjustbox}
\usepackage{subcaption}

\usepackage[hidelinks]{hyperref}
\usepackage{array}
\newcolumntype{x}[1]{>{\centering\arraybackslash\hspace{0pt}}p{#1}}
\usepackage{amssymb}% http://ctan.org/pkg/amssymb
\usepackage{pifont}% http://ctan.org/pkg/pifont
\newcommand{\cmark}{\ding{51}}%
\newcommand{\xmark}{\ding{55}}%
\usepackage[flushleft]{threeparttable}
\usepackage[inline]{enumitem}

\begin{document}
\title{Creating a Scholarly Knowledge Graph from Survey Article Tables}
\titlerunning{Creating a Scholarly Knowledge Graph from Survey Article Tables}
% If the paper title is too long for the running head, you can set
% an abbreviated paper title here
%

\author{Allard Oelen\inst{1,2}\orcidID{0000-0001-9924-9153}\and
Markus Stocker\inst{2}\orcidID{0000-0001-5492-3212} \and
S\"oren Auer\inst{2,1}\orcidID{0000-0002-0698-2864}}

\institute{L3S Research Center, Leibniz University of Hannover, Germany \\
\email{oelen@l3s.de} \and 
TIB Leibniz Information Centre for Science and Technology, Germany
\email{\{markus.stocker,auer\}@tib.eu}} 

\authorrunning{Oelen et al.}
\maketitle              % typeset the header of the contribution
\begin{abstract}
Due to the lack of structure, scholarly knowledge remains hardly accessible for machines. Scholarly knowledge graphs have been proposed as a solution. Creating such a knowledge graph requires manual effort and domain experts, and is therefore time-consuming and cumbersome. In this work, we present a human-in-the-loop methodology used to build a scholarly knowledge graph leveraging literature survey articles. Survey articles often contain manually curated and high-quality tabular information that summarizes findings published in the scientific literature. Consequently, survey articles are an excellent resource for generating a scholarly knowledge graph. The presented methodology consists of five steps, in which tables and references are extracted from PDF articles, tables are formatted and finally ingested into the knowledge graph. To evaluate the methodology, 92 survey articles, containing 160 survey tables, have been imported in the graph. In total, $2\,626$ papers have been added to the knowledge graph using the presented methodology. The results demonstrate the feasibility of our approach, but also indicate that manual effort is required and thus underscore the important role of human experts. 

\end{abstract}
\section{Introduction}
Scholarly communication is mainly document-based and the communicated scholarly knowledge therefore hardly machine-actionable~\cite{mons2009nano}. Scholarly knowledge graphs have the potential to solve these issues by making knowledge structured and thus more machine processable. Existing initiatives for scholarly information systems, e.g., the Microsoft Academic Graph~\cite{herrmannova2016analysis} or Crossref~\cite{Lammey2015} mainly focus on bibliographic metadata and not on the actual research contributions. The Open Research Knowledge Graph (ORKG)~\cite{Jaradeh2019} aims to build a knowledge graph infrastructure that publishes the research contributions of scholarly publications rather than only the metadata. The approach is to crowdsource structured paper descriptions by including paper authors and domain experts. ORKG primarily relies on synergistically combining crowdsourcing and automated extraction rather than, as other systems such as Semantic Scholar\footnote{\url{https://www.semanticscholar.org}}, exclusively on automated techniques to extract knowledge from scholarly articles. Mainly because automated extraction methods, for example Natural Language Processing (NLP), do not have sufficient accuracy to generate the high-quality knowledge graph needed to obtain suitable state-of-the-art overviews for researchers. 

\begin{figure}[t]
    \centering
    \includegraphics[width=0.9\textwidth]{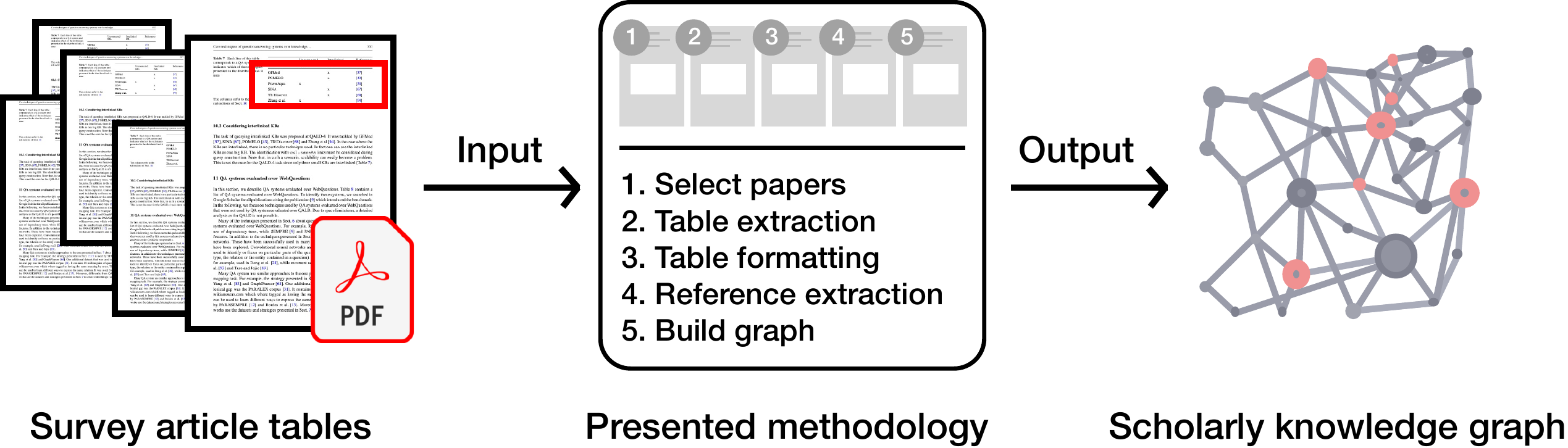}
    \caption{Systematic workflow in which survey articles are used to build a scholarly knowledge graph. The input of our methodology is survey articles in PDF format and the output is a scholarly knowledge graph.
    }
    \label{fig:paper-overview}
\end{figure}

In this work, we present a human-in-the-loop methodology to create a scholarly knowledge graph by extracting knowledge from survey tables. We leverage survey tables from literature review papers, specifically. Tables in survey papers generally consist of high-quality research data that has been manually curated by domain experts. Conducting a literature review is a labour-intensive task and writing a review article is often more time-consuming than writing a research article~\cite{Webster2002}. Compared to natural text, tables present information in a semi-structured manner, making the creation of a structured graph from such data less complicated. Additionally, survey tables present relevant information which is why the survey was conducted and published in the first place. We present a supervised approach to firstly extract data from survey articles and afterwards build a knowledge graph from this data. Compared to sole crowdsourcing, the approach of extracting knowledge is more efficient because the review has already been conducted by the authors of the survey paper. Taking into account the previously mentioned considerations, our work addresses the following research question: \textit{How to efficiently populate a scholarly knowledge graph with high-quality knowledge?} We propose a methodology for extracting tabular survey data. This methodology is used to create a scholarly knowledge graph from survey articles. An overview of the systematic workflow is depicted in Figure \ref{fig:paper-overview}.

The rest of this paper is structured as follows. Section~\ref{section:related-work} discusses the related work. Section~\ref{section:methodology} introduces the proposed five-step methodology for building the knowledge graph from survey articles. Section~\ref{section:results} presents the results. Section~\ref{section:discussion} discusses the present and future work. Finally, Section~\ref{section:conclusion} concludes the presented work.

\section{Related Work}
\label{section:related-work}
%Survey articles list and review scientific literature in specific domains or fields. 
Survey articles provide well-structured overviews of the literature~\cite{Wee2016}. The terms ``literature review'' and ``literature survey'' are sometimes used interchangeably in the literature, but we make the following distinction. We refer to the tables within review articles as \textit{literature surveys}. Together with a (textual) analysis and explanation, they form the \textit{literature review}. Among other things, literature reviews are helpful in delimiting the research problem, avoiding fruitless approaches~\cite{gall8c} and to discover new research directions~\cite{Hart1998}. Conducting a literature review is a complicated and time-consuming activity~\cite{Wee2016}. When literature reviews are not available for certain fields, its development could be weakened~\cite{Webster2002}. Because of the importance of literature surveys to scientific research, leveraging surveys to build a graph results in a high-quality and relevant scholarly knowledge graph. Some existing work with respect to semantifying literature surveys exists~\cite{10.1007/978-3-319-67008-9_25,Vahdati2019,Oelen2020}. However, those approaches are not (semi-)automated and are therefore not scaling well to larger amounts of survey articles. 

One aspect of the proposed methodology is table extraction from survey articles. Portable Document Format (PDF) is the most common format for scientific articles~\cite{Klampfl2014}. Extracting tables from PDF documents is a cumbersome process since the tabular structure is not stored within the file itself~\cite{Jiang2009a}. This means that regular PDF extraction tools are only able to extract the text within a table, but loosing the tabular structure. Tools that specifically focus on table extraction from PDF files use segmentation techniques to estimate the position of rows and columns~\cite{Hassan2007}. Corrêa et al. did a literature survey on table extraction tools~\cite{Correa2017}. They concluded that Tabula\footnote{\url{https://tabula.technology}} is the most suitable open-source tool. Based of these findings, we decided to use Tabula. Tabula is criticized because of the lack of documentation~\cite{Rosen1363917}, but for our use case this is not considered problematic. Another aspect of the proposed methodology is reference extraction from PDFs. Since every individual article referenced within a survey table is imported, metadata from this article should be collected. This is done by parsing the references that are used within a table. For this, we use the state-of-the-art PDF extraction tool GROBID~\cite{10.1007/978-3-319-67162-8_15}. GROBID focuses specifically on extracting bibliographic data from scholarly articles~\cite{Lopez2009}. Lipinski et al.~\cite{Lipinski2013a} compared GROBID to other PDF metadata extraction tools, and found out that GROBID performed best.

Publishing data as structured or semantic data is a well researched topic among various domains. For example, challenges related to publishing semantic open government data are similar to the challenges in our research. This includes extracting data from legacy documents, often in PDF format~\cite{Correa2014,Correa2017}. Furthermore, in the literature use cases are described on publishing unstructured data as semantic data (e.g.,~\cite{hyvonen2012publishing,Makela2012,Skjveland2013}). These existing approaches differ from our approach since they generally aim to semantify a homogeneous set of documents. This enables them to create data specific ontologies. In our case, this is not feasible since we work with a highly heterogeneous set of survey tables coming from different domains and comparing different aspects of papers. Table \ref{table:related-work-comparison} provides a related work overview. In this overview, our proposed method is compared to other related approaches. To the best of our knowledge, this work is the first to build a knowledge graph at scale from survey tables.

%We created a related work overview in the ORKG that compares our approach to other existing approaches.\footnote{\url{https://www.orkg.org/orkg/comparison/R36099}} 

% \begin{itemize}
% %\item About the usefulness of survey tables (e.g., they are providing important data etc.) 
% \item PDF extraction (grobid, tabula), accuracy and problems (first describe the struggles with PDF extraction, survey on tools (and why tabula was chosen), problems with reference extraction etc.) 
% \item CSVs for building knowledge graphs 
% \item (OpenRefine)
% \item Other approaches of building knowledge graphs from other datasets 
% \item About the usability of tables (A Tabular Survey of Automated Table Processing)
% \item About extracting scientific table contents (Unleashing Tabular Content to Open Data: A Survey on PDF Table Extraction Methods and Tools)
% \end{itemize}

\subsubsection{Use Case: Open Research Knowledge Graph.}
%- Discuss the previous work on making FAIR literature reviews (JCDL), and how the data that is collected in this research is suitable for the ORKG  
%As mentioned in the introduction, the extracted survey data is imported in the Open Research Knowledge Graph (ORKG). %The data could be imported in an arbitrary (scholarly) knowledge graph, such as the Microsoft Academic Graph or WikiData. However, 

\begin{table}[t]
\caption{Related work compared to the method presented in our study. The full comparison is available via the ORKG.\protect \footnotemark}
\centering 
\begin{adjustbox}{scale=.82}
{\setlength{\extrarowheight}{2pt}
\begin{tabular}{x{0.8cm}|c|x{2.5cm}|x{3cm}|x{1.2cm}|x{2cm}|x{0.5cm}|x{0.5cm}|x{0.5cm}} 
\toprule
\rotatebox[origin=c]{90}{\textbf{Study}} & \rotatebox[origin=c]{90}{\textbf{Name}} &
\rotatebox[origin=c]{90}{\textbf{Method automation}} &
\rotatebox[origin=c]{90}{\textbf{Scope}} & 
\rotatebox[origin=c]{90}{\textbf{Input format}} &
\rotatebox[origin=c]{90}{\textbf{Output format}} &
\rotatebox[origin=c]{90}{\textbf{KG\textsuperscript{a} creation}} &
\rotatebox[origin=c]{90}{\textbf{Reference extraction}} &
\rotatebox[origin=c]{90}{\textbf{User interface}} \\ \midrule 
This study & ORKG & Semi-Automatic & Survey tables & PDF & JSON, RDF\textsuperscript{b} & \cmark & \cmark & \cmark  \\[3ex] \hline 
\cite{R36089} & SemAnn & Semi-Automatic & Scholarly articles  & PDF & RDF\textsuperscript{b} & \xmark & \xmark & \cmark\\[1.0ex] \hline 
\cite{R36091} & Web Tables & Automatic & Web tables & HTML & JSON & \xmark & \xmark & \xmark \\[1.0ex] \hline 
\cite{R36093} & TableSeer & Automatic & Scholarly articles & PDF & Relational database & \xmark & \xmark & \cmark \\[3ex] \hline 
\cite{R36095} & TEXUS & Automatic & Documents (application agnostic) & PDF & Abstract table representation & \xmark & \xmark & \xmark \\[6ex] \hline 
\cite{R36097} & \textit{None} & Automatic & Web tables & HTML, Spreadsheet & Relational schema & \xmark & \xmark & \xmark \\[6.0ex]
\bottomrule

\multicolumn{4}{l}{%
  \begin{minipage}{8.5cm}%
      \small \textsuperscript{a} Knowledge Graph;
  \textsuperscript{b} Resource Description Framework
  \end{minipage}%
}
\end{tabular}}
\end{adjustbox}
\label{table:related-work-comparison}
\end{table}
\footnotetext{\url{https://www.orkg.org/orkg/comparison/R36099}}

Extracted survey data can be imported in a variety of different (scholarly) knowledge graphs, such as the Microsoft Academic Graph, Wikidata~\cite{Krotzsch2014} or ORKG. We chose ORKG as our use case for the following reasons. The ORKG provides tools that specifically focus on building paper comparisons (i.e., literature surveys), making it the most suitable infrastructure for this study. By using the extracted survey data, the ORKG automatically generates a similar tabular survey view as was originally presented in the review paper~\cite{Oelen:Sciknow19}. Additionally, the literature surveys within ORKG are compliant~\cite{Oelen2020} with the FAIR data principles~\cite{Wilkinson2016} thus making them Findable, Accessible, Interoperable and Reusable. The imported survey tables are FAIR in contrast to the originally presented ones in the non-FAIR PDF article. This has several benefits, among others: 
\begin{itemize}
    \item Comparisons can evolve over time, are not static and do not become stale after publication.
    \item Comparisons do represent a broader community consensus, since many researchers and curators can revise, discuss and annotate.
    \item Via the ORKG search interface it is possible to search for specific comparisons and to create dynamic custom comparison views. 
    \item Survey data can be reused by other researchers more easily because of its machine readable export formats (e.g., export as CSV or RDF).%\footnote{Resource Description Framework}).
\end{itemize}

\section{Methodology}
\label{section:methodology}

\begin{figure}[t]
    \centering
    \includegraphics[width=\textwidth]{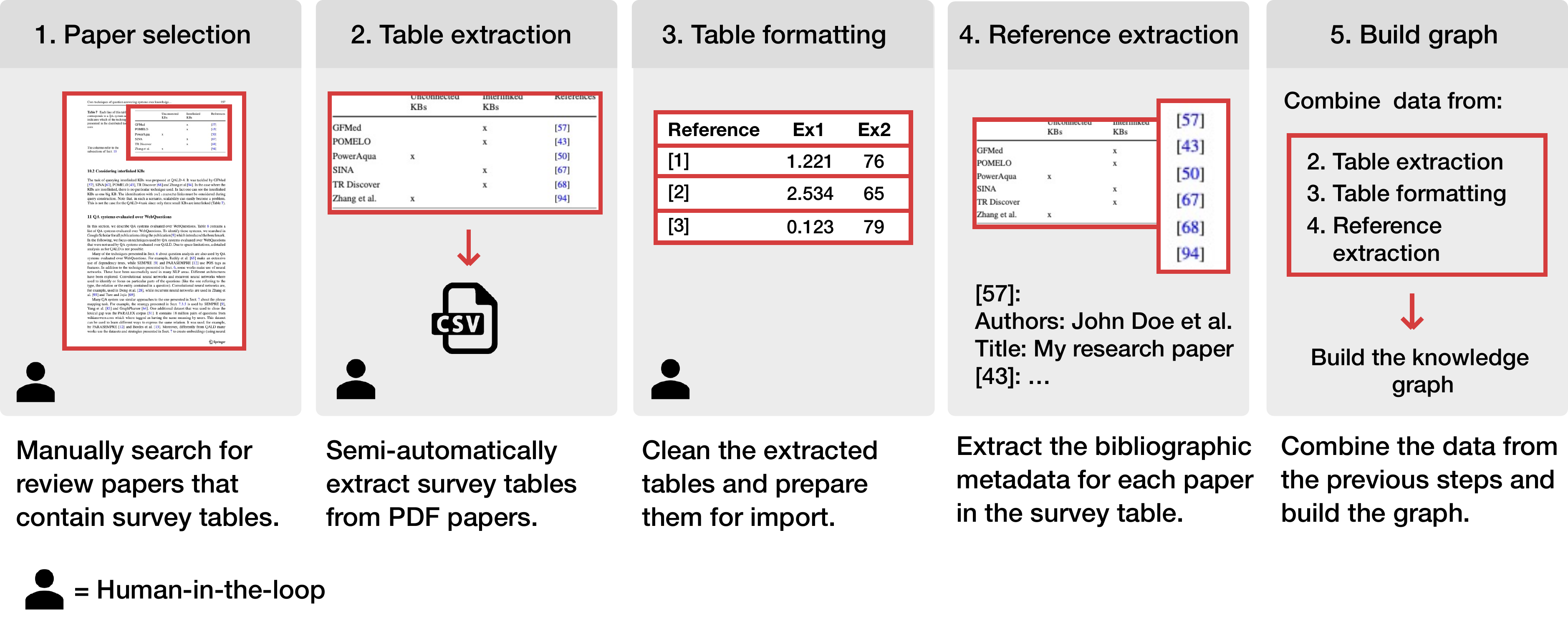}
    \caption{Methodology for importing survey tables into the scholarly knowledge graph.}
    \label{fig:methodology}
\end{figure}

We now present a five-step methodology for the creation of a scholarly knowledge graph from survey tables. In order to reach sufficient quality, the methodology takes a human-in-the-loop approach in which multiple steps require human interaction. Data quality improves with human evaluation and, if needed, correction of the extracted data. The methodology is displayed in Figure~\ref{fig:methodology}. The scripts required to perform the steps are available online.\footnote{\url{https://doi.org/10.5281/zenodo.3739427}}

\subsection{Paper Selection}
In the first step, suitable survey papers are selected based on multiple criteria. %A human is required to find suitable papers across multiple sources. 
The purpose is to find survey papers from a diverse range of domains. Therefore, a protocol has been designed to determine which papers are suitable for data extraction. The structured nature of the selection process is needed to be able to make conclusions about the percentage of survey papers that present the information in such a way that extracting data is relatively straightforward. 

\subsubsection{Search Strategy.}
Table~\ref{table:search-sources} lists the search engines used to find survey articles. Google Scholar is chosen to ensure that survey papers from various fields are searched. Additionally, ACM Digital Library has been selected because the ORKG currently focuses mainly on the Computer Science domain. The search is limited to 100 papers that are suitable for import. The following search criteria are used:
\begin{itemize}
    \item Google Scholar: the article title contains the term ``literature survey".
    \item ACM Digital Library:  queries ``literature review" and ``literature survey".
    \item The survey article has been published after 2002.
    \item The results are sorted by relevance.
\end{itemize}

\begin{table}[tb]
\centering
\caption{Search engines used to find survey articles.}
\label{table:search-sources}
\begin{adjustbox}{scale=.95}
\begin{tabular}{l|l|r}
\toprule
\textbf{Search engine}     & \textbf{Field} & 
\textbf{Evaluated papers} \\ \midrule
Google Scholar      & All & 335              \\ 
ACM Digital Library & Computer Science & 80 \\
\bottomrule
\end{tabular}
\end{adjustbox}
\end{table}

\noindent The rationale for selecting papers published after the year 2002 is because in general more recent papers are more interesting for research and should therefore have more priority in the scholarly knowledge graph. In the end, articles published before 2002 can still be part of the graph, since this criterion only applies to the survey articles themselves, and not to the papers being reviewed in those articles. 

\subsubsection{Selection Criteria.} 
\label{section:selection-criteria}
Papers that satisfy the inclusion criteria are selected for the import process. The inclusion criteria are defined as follows: 

\begin{enumerate}
    \item The article contains at least one table that lists scientific literature (i.e., the literature is presented in a semi-structured manner).
    \item The article compares literature based on published results and does not solely textually summarize the content of original papers. 
    \item The survey table should be in markup format and not included as raster image. %The table listing the literature should be included as text in the PDF (i.e. table that are included as image are discarded) 
    \item The table structure should be suitable for import (e.g., one table row should provide information about one publication).
    \item The article is written in English.
    %\item The article is accessible from within the network
\end{enumerate}

\noindent Inclusion criterion 1 ensures that a survey article does not only textually summarize the literature, but does also provide a semi-structured comparison (in tabular form). Although papers that are textually reviewing scientific literature are interesting for importing as well, it is out of scope for this work. Criterion 2 ensures only surveys that compare actual paper results are included. This excludes surveys researching, for instance, the growth of a field. Criterion 3 excludes tables in image format. This is because of the tabular extraction method we use, which is based on character extraction and does not use Optical Character Recognition (OCR) needed to support image extraction~\cite{Vasileiadis2017}. Criterion 4 only selects tables that are suitable for import. Our methodology does only support paper import when one row in a table represents one paper. Although minor changes can be made manually (e.g. merging multiple tables), in case the structure of the table deviates significantly from the required format, the table is excluded. Finally, criterion 5 ensures a homogeneous semantic integration into the currently English monolingual knowledge graph. The result of this step is a set of the selected papers in PDF format.  

\subsection{Table Extraction}
This step focuses on extracting the tables from the PDF files collected in the previous step. Not only the text within the table should be extracted, but the tabular structure should be preserved as well. %A commonly used and well performing tool~\cite{Lipinski2013a} for extracting data from scholarly articles is GROBID. However, this tool only provides the ability to extract text which means that the tabular structure gets lost. The PDF table extraction tool Tabula is evaluated as the best open-source tool to perform this task~\cite{Correa2017}.
As explained in the related work section, we use Tabula to perform the table extraction. Each PDF article is uploaded via the Tabula user interface. Afterwards, the regions of the tables are manually selected within the interface. Although Tabula provides a functionality to automatically detect tables, the accuracy is not sufficient for our use case. The performance is especially low for articles with a two-column layout. Additionally, not all tables within an article have to be extracted since not all of them are listing and comparing literature. Arguably, the manual selection method is most useful in this methodology since human judgment is needed in the selection process. Part of the extraction step is quality assurance after the extraction. When needed, extraction errors are manually fixed. Tabula supports two types of extraction, namely ``Stream'' and ``Lattice''. The Stream extraction method is based on white space between columns while Lattice is based on boundary lines between columns. During the extraction it is possible to switch between the different methods, which allows for selecting the best method for a particular table. The result of this step is a set of CSV files, in which each file represents one survey table from a review article. 

\subsection{Table Formatting}
The CSV files containing the extracted tables from the review articles should be formatted in a structure that is suitable for building a graph. Since the data from the CSV file is extracted automatically, all tables should have the same format. In this step, the formatting of the tables is changed when necessary. For some tables, a considerable amount of changes is required while for other tables only minor changes are needed. Changes could include merging, splitting, adding and removing both columns and rows. We use OpenRefine~\cite{verborgh2013using} to perform bulk operations on tables. A table is formatted in such a way that it adheres to the following rules:

\begin{enumerate}
    \item The first row of the table is the header.
    \item Each row represents one reviewed paper.
    \item Each row has a column called: ``Reference".
    \item The reference cell should contain the citation key for a paper .
    \item Non-literal values are prefixed with ``[R]" in the column header.
    \item When needed, abbreviations are replaced by the full value from the legend.
\end{enumerate}

\noindent For rule 2, in some cases a multidimensional table has to be flattened. This can often be accomplished by adding additional columns to the table. Also, in some cases a table has to be transposed to ensure that each row contains one paper. Rules 3 and 4 ensure that bibliographic metadata can be fetched for each paper in the next step. Rule 5 makes a distinction between literal values and resources. The default cell type is a literal, and when [R] is prefixed to a header label, the cells are considered as resources. Finally, rule 6 makes the content of the table readable without requiring the original text from the legend. Often table legends are used to condense information to improve user readability. 

\subsection{Extracting References}
As mentioned earlier, each table row represents one paper. For each row, there is a value that contains the reference key from the original paper. The reference key is often a numerical reference, in the form of [\textit{n}], where \textit{n} represents the reference number. In another frequently used citation style, the author names combined with their publication year is used as a reference key. The citation key is used to automatically capture the bibliographic metadata for an article. In order to extract references from article, we use the PDF extraction tool GROBID. GROBID processes the full PDF article. In the first place to extract all citations from the paper's reference list and then to connect the citation keys used in the text to their respective citation string. In case a reference key cannot be extracted from the paper's text, a reference key is generated automatically based on the author's name and publication year. %In order to improve the recall of the citation matching, matching citation keys was done with an edit distance~\cite{ristad1998learning} of four. This matches citation keys in case they are not identical, which may happen if letters get lost during the extraction step.

When the citation is extracted and parsed, five additional columns are appended to the table: paper title, authors, publication month, publication year and the DOI\footnote{Digital Object Identifier}. In case a citation key could not be automatically mapped to an actual citation, a citation can be provided manually. The full citation text can be copied directly from the paper (including paper title, authors etc.) and is then parsed by GROBID to get structured bibliographic metadata. To perform the process of adding references, we created a Python script.\footnote{File \texttt{4\_reference\_extraction.py} from \url{https://doi.org/10.5281/zenodo.3739427}} This script first tries to automatically fetch the metadata. In case the reference is not found, a command line input field is displayed to enter the citation manually. 

\subsection{Build Graph}
\begin{figure}[tb]
    \centering
    \includegraphics[width=0.95\textwidth]{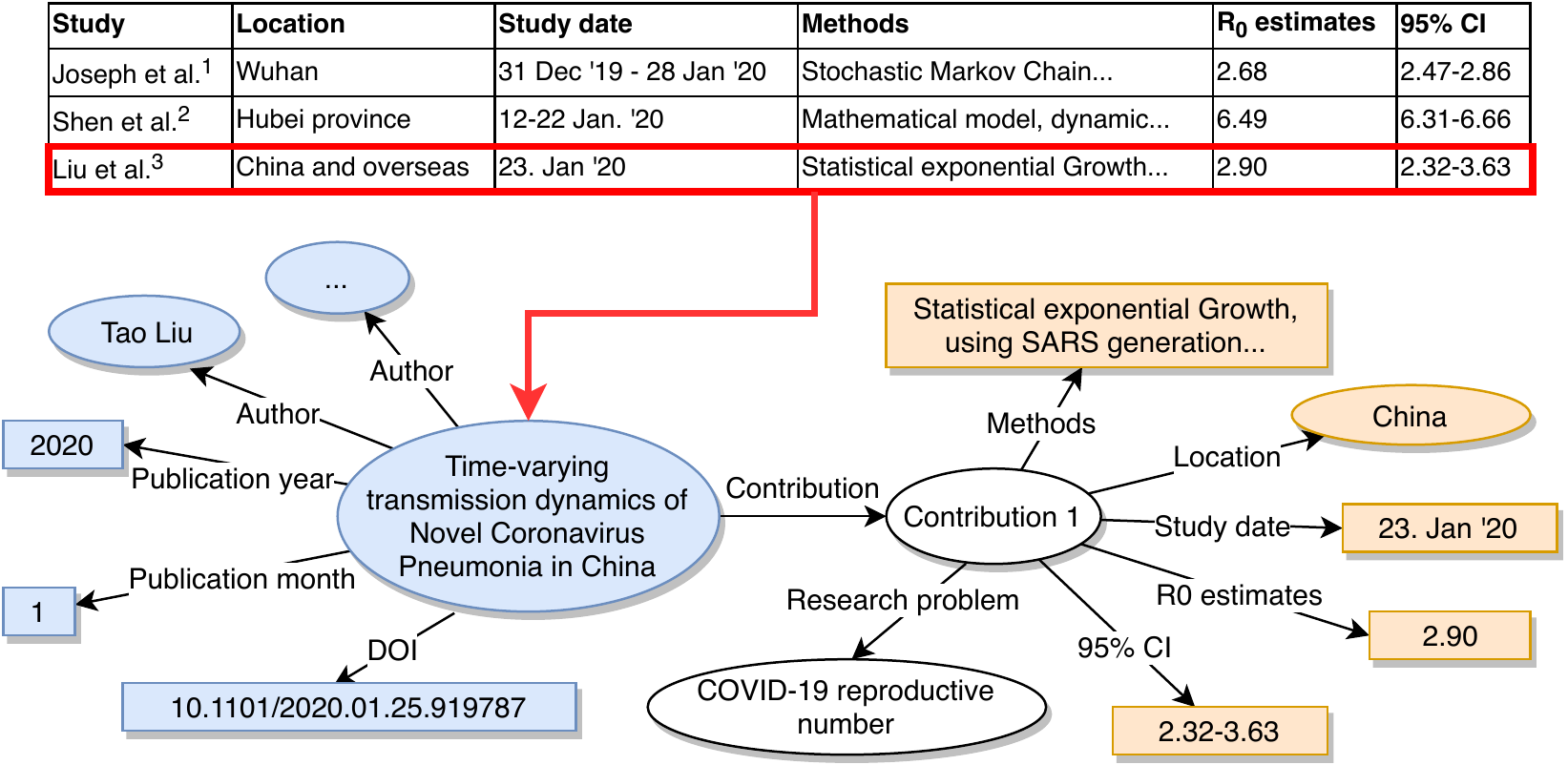}
    \caption{Example of the resulting subgraph for importing a single paper from a survey table. Metadata captured by reference extraction is displayed in blue. Data coming from the survey table is displayed in orange and ORKG specific data is displayed in white.}
    \label{fig:graph-paper-example}
\end{figure}

The final step is to build a knowledge graph from the previously created CSV files. An example of the resulting graph for a single paper is depicted in Figure~\ref{fig:graph-paper-example}. Firstly, a settings file is created which lists the table numbers, a suitable title for the table and a reference to the original survey article. The reference is required to attribute the work done by the authors of the survey article. The table title is manually created based on the original table caption. In case no suitable caption is available, a more suitable title is written. 

Next, a Python script\footnote{File \texttt{5\_build\_graph.py} from \url{https://doi.org/10.5281/zenodo.3739427}} is used to select all rows from the tables. For each row, a paper is added to the graph via the ORKG API. For each table, a comparison is created in ORKG. The title and reference from the previously generated settings file are attached to this comparison. The comparison can be used later in ORKG to generate the same tabular literature overview as originally presented in the survey paper. 

\subsection{User Interface}
\label{section:user-interface}
\begin{figure}[t]
  \begin{subfigure}{0.45\textwidth}
    \includegraphics[width=\linewidth]{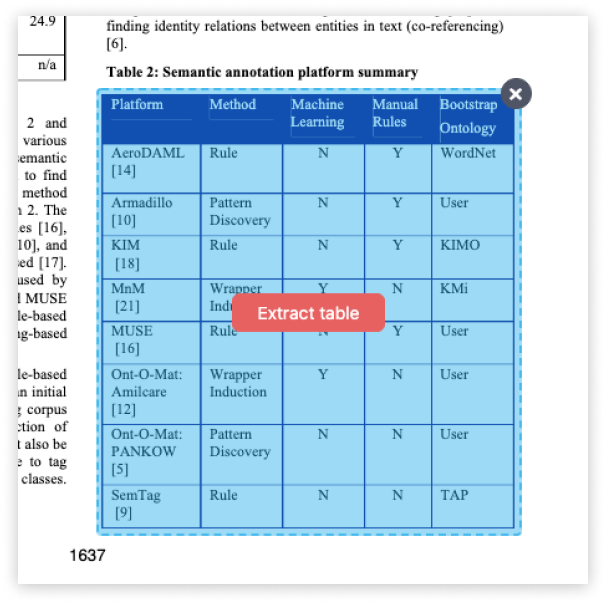}
    \caption{Select and extract table from PDF.} \label{figure:prototype-interface-1a}
  \end{subfigure}
  \hspace*{\fill}
  \begin{subfigure}{0.53\textwidth}
    \includegraphics[width=\linewidth]{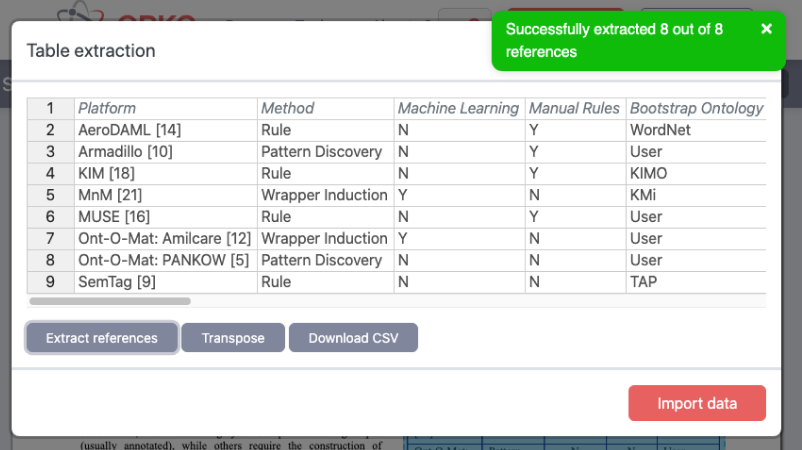}
    \caption{Fix table formatting, add references and ingest in graph.} \label{figure:prototype-interface-1b}
  \end{subfigure}
  \hspace*{\fill}
\caption{Survey table import User Interface integrating all steps from the methodology.} \label{figure:prototype-interface}
\end{figure}

Based on the steps from our methodology, a web User Interace (UI) is created that integrates all steps into a single interface. The interface provides a streamlined process for importing survey tables as depicted in Figure~\ref{figure:prototype-interface}.
%The steps of the methodology are integrated in a web User Interface (UI) to provide a streamlined process for importing survey tables. 
The UI is specifically designed to make importing a table an effortless task without the need of downloading any tools or the need to be able to operate these tools. In the background, the same tools from the methodology are used to extract tables (Tabula) and extract references (GROBID). The first step is to upload a PDF file and select the survey table within this file. Afterwards, the table is extracted and the formatting can be fixed with an integrated spreadsheet editor. Then, for each row the respective paper reference is extracted. Finally, the data is ingested in the knowledge graph. 

The UI is not used to import the surveys tables presented in Section~\ref{section:results}. The interface is designed to import individual survey tables rather than importing large amounts of tables at once. In the UI, all steps required to import a single table should be performed consecutively. To increase efficiency when importing large amounts of tables, it helps to first finish a step for all papers before moving to the next step. The UI provides a method to extend the graph beyond the extracted surveys from this work. In the future, this interface will therefore be integrated in the ORKG.

\section{Results}
\label{section:results}
In this section, we report the results of the import process for each step of the methodology. Table~\ref{table:results-summary} summarizes the results for all steps. 

\begin{table}
\centering
\caption{Summary of the results of all steps.}
\label{table:results-summary}
\begin{adjustbox}{scale=.95}

\begin{tabular}{lr}
\toprule
\multicolumn{1}{l|}{\textbf{Description}} & \textbf{Amount} \\ \midrule
\multicolumn{2}{l}{\textit{Paper selection}} \\ \midrule 
\multicolumn{1}{l|}{Amount of evaluated papers} & 415 \\ 
\multicolumn{1}{l|}{Amount of selected papers} & 92 \\ \midrule
\multicolumn{2}{l}{\textit{Table extraction}} \\ \midrule
\multicolumn{1}{l|}{Total amount of extractions (partial tables)} & 265 \\ 
\multicolumn{1}{l|}{Amount of extracted complete tables} & 160 \\ \midrule
\multicolumn{2}{l}{\textit{Reference extraction}} \\ \midrule
\multicolumn{1}{l|}{Found references} & $2\,069$ \\ 
\multicolumn{1}{l|}{Not found references} & $1\,137$ \\ \midrule
\multicolumn{2}{l}{\textit{Build graph}} \\ \midrule
\multicolumn{1}{l|}{Individual amount of imported papers} & $2\,626$ \\ \multicolumn{1}{l|}{Imported data cells (with metadata)} & $40\,584$ \\
\multicolumn{1}{l|}{Imported data cells (without metadata)} & $21\,240$ \\ 
 \bottomrule
\end{tabular}
\end{adjustbox}
\end{table}

\subsection{Paper Selection}
%The selected survey articles are published online
The dataset of the results are published online~\cite{oelenZenodoDataset}. This set contains the selected papers, the ORKG comparisons and the ingested papers. The selected papers file lists IDs, paper titles, table references, sources and references. The IDs are used to record any additional information about the import process for this specific paper. IDs are missing for papers that were selected in the first place, but were excluded after revising the inclusion criteria. Additionally, table references refer to the original table references used in the survey article. 

In total, 335 papers from Google Scholar were evaluated against the selection criteria described in Section~\ref{section:selection-criteria}. Out of these papers, 78 met the criteria and have therefore been selected for importing. From the ACM Digital Library 80 papers were evaluated and 14 papers have been selected. In total 22\% of the evaluated review papers are suitable to be imported with the presented approach.

\begin{table}[b]
\centering
\caption{Issues that occurred during the extraction of tables from the survey articles. Issues are counted per article. }
\label{table:extraction-issues}
\begin{adjustbox}{scale=.92}

\begin{tabular}{l|l|r}
\toprule
\textbf{\#} & \textbf{Issue}                                                    & \textbf{Percentage \%} \\ \midrule
1           & Columns are not extracted correctly                               & 26                  \\
2           & Rows are not extracted correctly                                  & 14                  \\ 
3           & Empty columns in the extracted table                              & 14                  \\ 
4           & Text not correctly recognized (e.g., missing letters or formulas) & 12                  \\ 
5           & Issue with table header text                                      & 12                  \\ 
6           & Vertical text not imported correctly                              & 4                   \\ 
7           & Cell value not supported (e.g., use of image instead of text check marks)          & 3                   \\ 
8           & Table within table not extracted correctly                        & 3      
\\
\bottomrule
\end{tabular}
\end{adjustbox}
\end{table}

\subsection{Table Extraction}
We extracted 160 tables from the 92 survey articles. In 22 cases, tables stretched across multiple pages, which results in a total of 265 extractions performed with Tabula. Table~\ref{table:extraction-issues} lists the most frequently occurred issues with the extraction. Issue 1 and 2 occur mostly when no boundary lines are present between table columns. In this case, the Stream extraction method has to be used, which often results in rows that are not correctly merged (e.g., multi-line sentences are put in separate rows while in the original table they are in the same row). Also, issue 3 is mostly present when using the Stream method. When the Lattice method can be used for the extraction, the result is generally of higher quality. When no table borders (or boundary lines) are present, this method does not work and the Stream method has to be used. Issue 4 is caused by general extraction errors, which can result in tables with wrongly extracted text. Additionally, formulas and other text styling are not supported, which compounds this issue. Issues 7 and 8 result in tables that are not, or only partially, imported. The other issues are self-explanatory. 

% \subsection{Table formatting}
% Todo

\subsection{Reference Extraction}

In total, we extracted unique $2\,626$ papers from $3\,206$ rows. For each paper, the respective citation was retrieved. In $2\,069$ cases the citation could be extracted automatically from the row (65\% of the cases). In $1\,137$ cases it was not possible to automatically extract the reference (35\% of the cases). For those cases, the citation is manually copied from the paper. There were multiple reasons why automatic reference extraction was not successful. Most issues occurred for references that used a numeric citation key. GROBID's performance for extracting numeric references from tables was low, oftentimes numeric table references were not recognized. The amount of rows is higher than the amount of extracted papers because multiple rows could refer to the same paper. Each paper only has one graph entry and any additional data is added to the existing paper. 

%Therefore, the references were used from other parts in the text. 

In case a reference is only used in a table and not somewhere else in the article, automatic reference extraction was oftentimes not possible. When an author name was used as citation key, problems occurred mostly because of the different citation styles. While some citation formats only use the last name of the first author, suffixed by \textit{et al.}, other formats could list all author names. When a format was used that deviates from the standard implementation, automatic extraction was not possible. 

\subsection{Build Graph}
In total, we added $2\,626$ papers to the knowledge graph. These papers are used in 160 different comparisons. A complete list of the generated ORKG comparisons and a list of all ingested papers is available via~\cite{oelenZenodoDataset}. In total, $21\,240$ table cells have been imported, excluding the bibliographic metadata. Including metadata, the total is $40\,584$ data cells.

\section{Discussion and Future Work}
\label{section:discussion}

%- Discuss how much time is takes on average to import a table (compared to doing this manually like we did before, maybe link to a python script of that process) 

\subsection{Time Performance}
%In Table~\ref{table:time-performace}, estimates of the time required to import a single table are displayed. The minimum amount of time was needed for a relatively small table, which was extracted without any issues. The maximum amount of time was needed for a table with a complex layout, stretched across multiple pages. Also, this table did not have boundary lines. In the end, most time was spend on fixing extraction issues. 
The presented methodology takes a human-in-the-loop approach as opposed to a fully automated approach. Compared to a fully manual approach, the proposed approach saves considerable time. In previous work~\cite{Oelen2020}, we manually imported only four survey articles. On average, this process took 4 hours per article. For each of the papers, a Python script was created specifically to import the survey table with its references and data. An example of such a script for one paper can be found online.\footnote{\url{https://gitlab.com/TIBHannover/orkg/orkg-papers/-/blob/master/question-answering-import.py}} For the methodology used in this paper, the time to import one survey article was on average 15 minutes. Compared to the 4 hours of the manual approach, this is considerably faster (i.e., 16 fold increase in speed). The minimum amount of time needed to import a relatively small table was 2 minutes. The table could be extracted without any issues. The maximum amount of required time was approximately 60 minutes. This was for a table with a complex layout, stretched across multiple pages. Also, this table did not have boundary lines. Most time was spend on fixing extraction issues. To further improve time performance, we identified two tasks that are time-consuming and can potentially be improved. The first task relates to fixing errors occurred during the table extraction by Tabula. Most errors occurred when tables did not have boundary lines between columns and rows. A potential solution, and possible future research direction, is to create an interface that supports manually drawing boundary lines between rows and columns. The second task is related to adding missing references, which have to be manually copied from the PDF article. In total, 65\% of the references were extracted automatically. By applying more advanced heuristics to match reference keys with their respective reference, this percentage can be improved.

% \begin{table}[t]
% \centering
% \caption{Time required to import a single survey table}
% \label{table:time-performace}
% \begin{adjustbox}{scale=.90}
% \begin{tabular}{l|l|l}
% \toprule
% \textbf{Minimum} & \textbf{Average} & \textbf{Maximum} \\ \midrule 
% 2 minutes & 15 minutes & 60 minutes \\
% \bottomrule
% \end{tabular}
% \end{adjustbox}
% \end{table}

%To further reduce the time, some steps of the methodology can be improved. For example, during the reference extraction, 65\% of the references were found automatically. By applying more advanced heuristics to match reference keys with their respective reference, this percentage can be improved. Additionally, the formatting step was time-consuming for some tables, especially when errors occurred during the extraction step. Most errors occurred when tables did not have any boundary lines between columns and rows. A potential solution, and possible future research direction, is to improve the extraction step. In the first place, Tabula can be improved by providing better support for two-column article layouts. In this case, tables could be detected automatically. Furthermore, to improve the extraction itself, a user could manually draw the boundary lines between columns of the table. Although this requires more manual labour in this step, the extracted data is of higher quality, which saves time later during the formatting step. 

\subsection{Impact of Methodology}
The impact of the methodology relates to the amount of survey papers that are suitable for our approach (i.e., surveys representing information in tabular format). To order to provide insights on the impact, a structured search protocol has been employed in the paper selection step. As the results show, out of the 415 evaluated papers, 92 of them are suitable to be imported. This indicates that since 2002, 22\% of the published survey papers contain comparison tables. Therefore, arguably our methodology can have considerable impact when applied more broadly. In the paper selection, non-survey papers were excluded. However, it is not uncommon for research articles to also contain tables with related work (e.g., Table \ref{table:related-work-comparison} in this article). Thus the paper selection step could be extended to also include other articles to have a broader impact. 

\subsection{Semantics of Data}
The extracted knowledge graph consists of structured scholarly data. The quality of the knowledge graph could be further improved by providing more semantics to the data. Currently, a primitive method is used to map existing properties and resources. This is based on a lookup by resource label, in case a result is found, the resource is mapped. If not, a new resource is created. A more advanced mapping of resources and properties to existing ontologies improves the machine readability of the data. Tables containing large amounts of natural text (e.g., textually describing a methodology) could be further processed using named entity recognition and linking. This results in more structured data and therefore a higher quality knowledge graph. Approaches to improve the overall quality of the graph are part of future work.

% Furthermore, the extracted data is imported without much additional data processing. Manually processing the data is an infeasible task due to the large amounts of data. However, automatically processing the data using named entity recognition and linking could further improve it. It is possible to automatically extract data from natural text using named entity recognition and linking.

% The following triple, in the form of \texttt{(subject, predicate, object)}, is of an imported table: \texttt{ (Paper,  Highlights, "The tool was released to the NWN community and downloaded 6,000 times")}. This natural text sentence could be replaced by more structured triples, e.g.:
% \\\\
% \texttt{(Paper, Highlights, \_Highlight)} \\
% \texttt{(\_Highlight, ReleasedTo, NwnCommunity)} \\
% \texttt{(\_Highlight, Downloads, "6000")}\\\\
% This makes the paper more machine readable and therefore better comparable. However, manually generating such descriptions is a time-consuming process and is out-of-scope for this research. 
% Apart from manually creating more semantic data, other methods, such as 

% Other methods to automatically get structured data from a natural text description are via named entity recognition and linking.  

\subsection{Future Research Directions}

In total, we extracted 92 survey articles from a variety of domains. In the future, more survey articles will be ingested in ORKG. This will be done for multiple domains. The User Interface (UI) presented in Section \ref{section:user-interface} can be used to support users to import survey tables. The UI will be further improved to make to process more efficient. Due to the dynamic nature of the interface (especially compared to a regular spreadsheet editor), mapping properties and resources to existing concepts is better supported. In the end, we aim to import as many surveys from a specific domain as possible. There are several reasons why such an approach is useful. In the first place, ORKG can serve as a digital library for literature surveys. As discussed in the related work, the platform provides tools to better find and organize surveys. Additionally, when all existing reviews for a domain are imported, the ORKG can be used as a source to find literature surveys. In case a survey is not present in the ORKG, it means that is does not (yet) exist. This can be used as a basis to start working on new literature surveys.
%The presented methodology can be used to also import non-survey tables from PDF articles. The methodology specifically focuses on survey tables to build a knowledge graph. However, with minor adjustments, the methodology could also be adopted for importing other types of scholarly tables. In this case, the \textit{Reference extraction} step can be skipped.

%Finally, the methodology could serve a starting point for fully automated knowledge graph extraction from survey tables. The performance of fully automated extraction relies on the accuracy of multiple components, \begin{enumerate*}[label=(\arabic*)] \item automatic table detection, \item reference extraction within tables and \item automatic table extraction.\end{enumerate*} This makes it a complex task since if a single component does not perform well, the overall performance will drop. 

\section{Conclusions}
\label{section:conclusion}
Knowledge graphs are useful to make scholarly knowledge more machine actionable. Manually building such a knowledge graph is time-consuming and requires the expertise of paper authors and domain experts. In order to efficiently build a high-quality scholarly knowledge graph, we leverage survey tables from review articles. Generally, survey tables contain high-quality, relevant, semi-structured and manually curated data, and are therefore an excellent source for building a scholarly knowledge graph. We presented a methodology used to extract $2\,626$ papers from 92 survey articles. The methodology adopts a human-in-the-loop approach to ensure the quality and usefulness of the extracted data. Compared to manually reviewing and entering research data, or to manually importing literature surveys, the methodology is considerably more efficient. In conclusion, the presented methodology provides a full pipeline that can be used to extract knowledge from PDF documents and represent the extracted knowledge in a knowledge graph. The corresponding evaluation with survey articles demonstrates the effectiveness and efficiency of the proposed methodology. 
\\\\
\noindent
\textbf{Acknowledgements.} This work was co-funded by the European Research Council for the project ScienceGRAPH (Grant agreement ID: 819536) and the TIB Leibniz Information Centre for Science and Technology. We want to thank our colleagues Mohamad Yaser Jaradeh and Kheir Eddine Farfar for their contributions to this work.

\bibliographystyle{splncs}
\bibliography{refs-manually} % 

\begin{thebibliography}{10}
\providecommand{\url}[1]{\texttt{#1}}
\providecommand{\urlprefix}{URL }
\providecommand{\doi}[1]{https://doi.org/#1}

\bibitem{R36097}
Adelfio, M.D., Samet, H.: Schema extraction for tabular data on the web.
  Proceedings of the VLDB Endowment  \textbf{6},  421--432 (2013).
  \doi{10.14778/2536336.2536343}

\bibitem{Correa2014}
Corr{\^{e}}a, A.S., Corr{\^{e}}a, P.L.P., Da~Silva, F.S.C.: {Transparency
  portals versus open government data. An assessment of openness in Brazilian
  municipalities}. ACM International Conference Proceeding Series pp. 178--185
  (2014). \doi{10.1145/2612733.2612760}

\bibitem{Correa2017}
Corr{\^{e}}a, A.S., Zander, P.O.: {Unleashing tabular content to open data: A
  survey on PDF table extraction methods and tools}. ACM International
  Conference Proceeding Series pp. 54--63 (2017). \doi{10.1145/3085228.3085278}

\bibitem{10.1007/978-3-319-67008-9_25}
Fathalla, S., Vahdati, S., Auer, S., Lange, C.: {Towards a Knowledge Graph
  Representing Research Findings by Semantifying Survey Articles}. In: Research
  and Advanced Technology for Digital Libraries. pp. 315--327 (2017)

\bibitem{gall8c}
Gall, M.D., Borg, W.R.: {Educational Research: An introduction (sixth
  edition)}. White Plains, NY: Longman Publishers USA  (1996)

\bibitem{Hart1998}
Hart, C.: {Doing a Literature Review: Releasing the Social Science Research
  Imagination}. Sage (1998)

\bibitem{Hassan2007}
Hassan, T., Baumgartner, R.: {Table recognition and understanding from PDF
  files}. Proceedings of the International Conference on Document Analysis and
  Recognition, ICDAR pp. 1143--1147 (2007). \doi{10.1109/ICDAR.2007.4377094}

\bibitem{herrmannova2016analysis}
Herrmannova, D., Knoth, P.: {An analysis of the microsoft academic graph}.
  D-lib Magazine  \textbf{22}(9/10) (2016).
  \doi{10.1045/september2016-herrmannova}

\bibitem{hyvonen2012publishing}
Hyv{\"{o}}nen, E.: {Publishing and using cultural heritage linked data on the
  semantic web}. Synthesis Lectures on the Semantic Web: Theory and Technology
  \textbf{2}(1),  1--159 (2012). \doi{10.2200/S00452ED1V01Y201210WBE003}

\bibitem{Jaradeh2019}
Jaradeh, M.Y., Oelen, A., Farfar, K.E., Prinz, M., D'Souza, J., Kismih{\'{o}}k,
  G., Stocker, M., Auer, S.: {Open research knowledge graph: Next generation
  infrastructure for semantic scholarly knowledge}. K-CAP 2019 - Proceedings of
  the 10th International Conference on Knowledge Capture pp. 243--246 (2019).
  \doi{10.1145/3360901.3364435}

\bibitem{Jiang2009a}
Jiang, D., Yang, X.: {Converting PDF to HTML approach based on text detection}.
  Proceedings of the 2nd international conference on interaction sciences:
  Information technology, culture and human  \textbf{403},  982--985 (2009).
  \doi{10.1145/1655925.1656103}

\bibitem{Klampfl2014}
Klampfl, S., Granitzer, M., Jack, K., Kern, R.: {Unsupervised document
  structure analysis of digital scientific articles}. International Journal on
  Digital Libraries  \textbf{14}(3-4),  83--99 (2014).
  \doi{10.1007/s00799-014-0115-1}

\bibitem{10.1007/978-3-319-67162-8_15}
K{\"{o}}rner, M., Ghavimi, B., Mayr, P., Hartmann, H., Staab, S.: {Evaluating
  Reference String Extraction Using Line-Based Conditional Random Fields: A
  Case Study with German Language Publications}. In: New Trends in Databases
  and Information Systems. pp. 137--145. Cham (2017)

\bibitem{Krotzsch2014}
Krotzsch, M., Vrandecic, D.: {Wikidata : A Free Collaborative Knowledge Base}.
  Communications of the ACM  \textbf{57}(10),  78--85 (2014).
  \doi{10.1145/2629489}

\bibitem{Lammey2015}
Lammey, R.: {CrossRef text and data mining services}. Insights: the UKSG
  Journal  \textbf{28}(2),  62--68 (2015). \doi{10.1629/uksg.233}

\bibitem{R36091}
Lehmberg, O., Ritze, D., Meusel, R., Bizer, C.: A large public corpus of web
  tables containing time and context metadata. Proceedings of the 25th
  International Conference Companion on World Wide Web - WWW '16 Companion
  (2016). \doi{10.1145/2872518.2889386}

\bibitem{Lipinski2013a}
Lipinski, M., Yao, K., Breitinger, C., Beel, J., Gipp, B.: {Evaluation of
  header metadata extraction approaches and tools for scientific PDF
  documents}. Proceedings of the ACM/IEEE Joint Conference on Digital Libraries
  pp. 385--386 (2013). \doi{10.1145/2467696.2467753}

\bibitem{R36093}
Liu, Y., Bai, K., Mitra, P., Giles, C.L.: Tableseer: automatic table metadata
  extraction and searching in digital libraries. Proceedings of the 2007
  conference on Digital libraries - JCDL '07  (2007).
  \doi{10.1145/1255175.1255193}

\bibitem{Lopez2009}
Lopez, P.: {GROBID: Combining automatic bibliographic data recognition and term
  extraction for scholarship publications}. International conference on theory
  and practice of digital libraries  \textbf{5714},  473--474 (2009).
  \doi{10.1007/978-3-642-04346-8\_62}

\bibitem{Makela2012}
M{\"{a}}kel{\"{a}}, E., Hyv{\"{o}}nen, E., Ruotsalo, T.: {How to deal with
  massively heterogeneous cultural heritage data - Lessons learned in
  CultureSampo}. Semantic Web  \textbf{3}(1),  85--109 (2012).
  \doi{10.3233/sw-2012-0049}

\bibitem{mons2009nano}
Mons, B., Velterop, J.: Nano-publication in the e-science era. In: Workshop on
  Semantic Web Applications in Scientific Discourse (SWASD 2009). pp. 14--15
  (2009)

\bibitem{Oelen:Sciknow19}
Oelen, A., Jaradeh, M.Y., Farfar, K.E., Stocker, M., Auer, S.: {Comparing
  Research Contributions in a Scholarly Knowledge Graph}. In: Proceedings of
  the Third International Workshop on Capturing Scientific Knowledge
  (SciKnow19). pp. 21--26 (2019)

\bibitem{Oelen2020}
Oelen, A., Jaradeh, M.Y., Stocker, M., Auer, S.: {Generate FAIR Literature
  Surveys with Scholarly Knowledge Graphs}. JCDL '20: The 20th ACM/IEEE Joint
  Conference on Digital Libraries (In Press)  (2020).
  \doi{10.1145/3383583.3398520}

\bibitem{oelenZenodoDataset}
Oelen, A., Stocker, M., Auer, S.: {Dataset for Creating a Scholarly Knowledge
  Graph from Survey Article Tables} (2020). \doi{10.5281/ZENODO.3735152}

\bibitem{R36095}
Rastan, R., Paik, H.Y., Shepherd, J.: Texus. Proceedings of the 2015 ACM
  Symposium on Document Engineering - DocEng '15  (2015).
  \doi{10.1145/2682571.2797069}

\bibitem{Rosen1363917}
Ros, G.: {Analysis of Tabula : A PDF-Table extraction tool} (2019)

\bibitem{Skjveland2013}
Skj{\ae}veland, M.G., Lian, E.H., Horrocks, I.: {Publishing the Norwegian
  Petroleum Directorate's FactPages as semantic web data}. International
  Semantic Web Conference  \textbf{8219},  162--177 (2013).
  \doi{10.1007/978-3-642-41338-4\_11}

\bibitem{R36089}
Takis, J., Islam, A.S., Lange, C., Auer, S.: Crowdsourced semantic annotation
  of scientific publications and tabular data in pdf. SEMANTICS '15 Proceedings
  of the 11th International Conference on Semantic Systems  (2015).
  \doi{10.1145/2814864.2814887}

\bibitem{Vahdati2019}
Vahdati, S., Fathalla, S., Auer, S., Lange, C., Vidal, M.E.: {Semantic
  Representation of Scientific Publications}. International Conference on
  Theory and Practice of Digital Libraries  \textbf{11799},  375--379 (2019).
  \doi{10.1007/978-3-030-30760-8\_37}

\bibitem{Vasileiadis2017}
Vasileiadis, M., Kaklanis, N., Votis, K., Tzovaras, D.: {Extraction of tabular
  data from document images}. Proceedings of the 14th Web for All Conference,
  W4A  (2017). \doi{10.1145/3058555.3058581}

\bibitem{verborgh2013using}
Verborgh, R., De~Wilde, M.: {Using OpenRefine}. Packt Publishing Ltd (2013)

\bibitem{Webster2002}
Webster, J., Watson, R.T.: {Analyzing the Past to Prepare for the Future:
  Writing a Literature Review.} MIS Quarterly  \textbf{26}(2),  xiii -- xxiii
  (2002)

\bibitem{Wee2016}
Wee, B.V., Banister, D.: {How to Write a Literature Review Paper?} Transport
  Reviews  \textbf{36}(2),  278--288 (2016).
  \doi{10.1080/01441647.2015.1065456}

\bibitem{Wilkinson2016}
Wilkinson, M.D., Dumontier, M., Aalbersberg, I.J., Appleton, G., Axton, M.,
  Baak, A., Blomberg, N., Boiten, J.W., da~Silva~Santos, L.B., Bourne, P.E.,
  Bouwman, J., Brookes, A.J., Clark, T., Crosas, M., Dillo, I., Dumon, O.,
  Edmunds, S., Evelo, C.T., Finkers, R., Gonzalez-Beltran, A., Gray, A.J.,
  Groth, P., Goble, C., Grethe, J.S., Heringa, J., t~Hoen, P.A., Hooft, R.,
  Kuhn, T., Kok, R., Kok, J., Lusher, S.J., Martone, M.E., Mons, A., Packer,
  A.L., Persson, B., Rocca-Serra, P., Roos, M., van Schaik, R., Sansone, S.A.,
  Schultes, E., Sengstag, T., Slater, T., Strawn, G., Swertz, M.A., Thompson,
  M., Van Der~Lei, J., Van~Mulligen, E., Velterop, J., Waagmeester, A.,
  Wittenburg, P., Wolstencroft, K., Zhao, J., Mons, B.: {Comment: The FAIR
  Guiding Principles for scientific data management and stewardship}.
  Scientific Data  \textbf{3}, ~1--9 (2016). \doi{10.1038/sdata.2016.18}

\end{thebibliography}

\end{document}